\documentclass{article}

\usepackage{arxiv}
\usepackage{amsmath}
\usepackage[utf8]{inputenc} 
\usepackage[T1]{fontenc}    
\usepackage{hyperref}       
\usepackage{url}            
\usepackage{booktabs}       
\usepackage{amsfonts}       
\usepackage{nicefrac}       
\usepackage{microtype}      
\usepackage{cleveref}       
\usepackage{graphicx}
\usepackage{natbib}
\usepackage{doi}

\usepackage{thumbpdf,lmodern}
\usepackage{framed}
\usepackage{mathtools}
\usepackage{caption}
\usepackage{subcaption}
\usepackage{multirow}

\usepackage{xspace,color}
\usepackage{listings}
\usepackage{courier}

\lstnewenvironment{Code}[1][]{\lstset{language=R}}{}
\newcommand{\code}[1]{\lstinline{#1}}  

\let\proglang=\textsf

\newcommand{\pkg}[1]{{\fontseries{m}\fontseries{b}\selectfont #1}}
\newcommand{\fct}[1]{{\fontseries{m}\fontseries{b}\selectfont #1()}}
\newcommand\uidsim{\stackrel{\mathclap{iid}}{\sim}}

\let\oldcite=\cite
\let\oldcitep=\citep 
\renewcommand{\cite}[1]{\textcolor[rgb]{0,0,1}{\oldcite{#1}}}
\renewcommand{\citep}[1]{\textcolor[rgb]{0,0,1}{\oldcitep{#1}}}

\hypersetup{
	linktocpage=true,
	pdfborder={0 0 0},
	pdfborderstyle={/S/U/W 0},
	hyperindex=true,
	bookmarks=true,
	bookmarksopen=true,
	bookmarksnumbered=true,
}

\title{remiod: Reference-based Controlled Multiple Imputation of Longitudinal Binary and Ordinal Outcomes with non-ignorable missingness}

\date{}

\author{ 
	Tony~Wang \\
	iMEDacs.com\\
	\texttt{xwang@imedacs.com} \\
	\And
	Ying~Liu \\
	iMEDacs.com\\
	\texttt{yingliu@imedacs.com} \\
}

\renewcommand{\headeright}{}
\renewcommand{\undertitle}{}
\renewcommand{\shorttitle}{\textit{remiod} for controlled multiple imputation}

\hypersetup{
pdftitle={remiod: Reference-based Controlled Multiple Imputation of Longitudinal Binary and Ordinal Outcomes with non-ignorable missingness},
pdfauthor={Tony~Wang, Ying~Liu},
pdfkeywords={controlled imputation, Bayesian, cumulative logistic model, generalized linear model, R, JAGS.},
}

\begin{document}
\maketitle

\noindent\makebox[\linewidth]{\rule{0.77\paperwidth}{0.4pt}}\\

\begin{abstract}
Missing data on response variables are common in clinical studies. Corresponding to the uncertainty of missing mechanism, theoretical frameworks on controlled imputation have been developed. In practice, it is recommended to conduct a statistically valid analysis under the primary assumptions on missing data, followed by sensitivity analysis under alternative assumptions to assess the robustness of results. Due to the availability of software, controlled multiple imputation (MI) procedures, including delta-based and reference-based approaches, have become popular for analyzing continuous variables under missing-not-at-random assumptions. Similar tools, however, still limit application of these methods to categorical data. In this paper, we introduce the R package \textbf{remiod}, which utilizes the Bayesian framework to perform imputation in regression models on binary and ordinal outcomes. Following outlining theoretical backgrounds, usage and features of \textbf{remiod} are described and illustrated using examples.
\end{abstract}

\noindent\makebox[\linewidth]{\rule{0.77\paperwidth}{0.4pt}}

\section[Introduction]{Introduction} \label{sec:intro}

Despite all efforts made to minimize missing data through careful planning and conduct, the amount of missing data could be nontrivial in most clinical studies. Missing data, especially on outcome variables, poses a challenge in data analysis of clinical studies since inappropriate handling of missing data could lead to biased and inefficient estimation of treatment effect. Regulatory guidelines recommend to conduct a statistically valid analysis under the primary assumptions on missing data, followed by sensitivity analysis under alternative assumptions to assess the robustness of results. 

Assuming all information about missingness are collected, the relationship between missing data and the values of variables in the data matrix can be grouped into three classes of mechanisms \citep{rubin2004}: missing completely at random (MCAR), where the probability of missingness is unrelated to observed or unobserved data; missing at random (MAR), where the probability of missingness is unrelated to unobserved data conditional on observed data; and missing not at random (MNAR), where the probability of missingness depends on unobserved data conditional on observed data. For likelihood-based analysis, MCAR and MAR implicate that the impact of missingness might be ignored, i.e. ignorable missingness. From a practical perspective, ignorability  means that the process (and its related parameters) that generates missing data can be ignored and we can focus on modeling the distribution of the hypothetical complete data. However, being “ignorable” does not mean that we can be entirely careless about missing values, such as removing all incomplete cases before doing the analysis (i.e., complete-case analysis) \citep{hezhang2021}. Another complexity is that MAR and MNAR cannot be distinguished from observed data. Thus, it is essential that the assumptions of analytic approaches are scientifically plausible and clearly stated.

Based on an assumption of missing mechanism, replacing missing data with substituted values, i.e. imputation, becomes popular with theoretical advancement in methodology for analyzing data with missingness. To incorporate the uncertainty of missing data, multiple values are imputed for each missing observation, producing multiple complete datasets. Following analysis of these datasets using standard complete data techniques, the multiple parameter estimates are pooled using Rubin’s rules \citep{rubin2004} to give a single estimate. This strategy separating imputation from analysis step disentangles the complex between missingness and estimation, thus making the challenge of dealing with missing data relatively straightforward. However, to ensure expected performance from analysis with multiple imputation (MI), the relations between the outcome and the covariates, which are modeled in the analysis step, must be included explicitly in these imputation models \citep{pmid16980150, pmid27042954, pmid28034175}

In the field of clinical studies, MAR assumption for missingness is often reasonable, thus being chosen as the basis of primary analysis in most studies \citep{national2010prevention}. Under MAR, the distribution of a participants’ data at the later stage of the study given their earlier observed data, does not depend on whether the data at the later stage of the study were observed. To assess the robustness of results based on MAR, most often sensitivity analyses based on plausible assumptions that depart from MAR should be performed. For analysis under MNAR, selection models, shared parameter models, or pattern mixture models (PMM)  can be employed through maximum likelihood, within a Bayesian framework, or multiple imputation \citep{9780470740521, 9780470849811}. Within the PMM framework, controlled MI is gaining more attention due to accessibility. Principled approaches of controlled MI includes:
\begin{itemize}
	\item Reference-based imputation: developed specifically for the setting of randomized clinical trials (RCT). In brief, the parameters of the MAR imputation model from a specified reference group (typically the placebo/control group) are borrowed to impute missing data in other groups in the trial in a contextually relevant manner \citep{pmid24138436, kenward2015controlled}. 
	\item delta-adjusted imputation: altering the MAR imputation distribution using a specified numerical sensitivity parameter $\delta$ \citep{carpenter2007missing, cro2020sensitivity} to explore the impact of a better or worse outcome for the unobserved, relative to that predicted based on the observed data under MAR.  
\end{itemize}

It is obvious that the controlled MI approach enables to explore the impact of contextually relevant assumptions, and makes them readily interpretable to clinical colleagues.

For continuous outcomes, computational algorithms on controlled MI have been investigated \citep{pmid24138436, Tang2015AnEM, pmid27010771, tang2017sas}. \proglang{SAS} macros for implementing reference-based MI on continuous outcomes are available \citep{fivemacro}. \proglang{STATA} has \pkg{mimix} command to implement the reference-based analysis for longitudinal continuous data \citep{pmid29398978}. In \proglang{R}, package \pkg{mlmi} implements a maximum likelihood MI version of reference-based imputation for repeatedly measured continuous endpoints. The controlled imputation methods have also been extended to time-to-event endpoints and recurrent events \citep{Keene2014, Tang2017recur}. For checking the assumption of independent censoring, package \pkg{InformativeCensoring} can be used. For recurrent event data, package \pkg{dejaVu} provides functions to performs reference-based MI proposed by \cite{Keene2014}.

For categorical outcomes, specifically for binary and ordinal responses, there still lack of software to facilitate the implementation of controlled imputation, although computational algorithm has been investigated \citep{pmid28073767, Tang2018sim}. In this paper, we introduce the \proglang{R} package \pkg{remiod}, which implements controlled MI for binary and ordinal outcomes.  Section 2 briefly describes the theoretical background of controlled MI methods for binary and ordinal data. The general structure of \pkg{remiod}\footnote{Available at \href{https://github.com/xsswang/remiod}{https://github.com/xsswang/remiod}} is outlined in Section 3, followed by an description of the example datasets that are used throughout the paper in Section 4. Details about model specification, settings controlling the MCMC sampling, and MI data extraction functions that can be applied after fitting the model are given in Sections 5. Application of \pkg{remiod} is illustrated through analyzing National Institute of Mental Health (NIMH) schizophrenia collaborative study. We conclude the paper with an outlook of planned extensions and discuss the limitations that are introduced by the assumptions made in the fully Bayesian approach.

\section{Theoretical background} \label{sec:theory}

\cite{Tang2018sim} proposed a sequential regression approach based on factorization of the joint distribution of longitudinal outcome:
\begin{equation} \label{eq:factorize}
f(y_{i1},...,y_{iJ}) \quad = \quad \prod_{j=1}^{s_i} f(y_{ij}|\mathbf{z}_{ij},\mathbf{\theta}_j) \prod_{j=s_i+1}^{J} g(y_{ij}|\mathbf{z}_{ij},\mathbf{\theta}_j, \tilde{\mathbf{\theta}}_j),
\end{equation}
where $\mathbf{y}_i = (y_{i1},..., y_{iJ})'$ be the outcomes of participant $i (i=1,...,n)$ having $J$ $(j=1,...,J)$ post-baseline visits, and $\mathbf{z}_i = (\mathbf{x}'_i, \mathbf{y}'_{i,j-1} )$, $\mathbf{z}_{i1} =\mathbf{x}_i = (x_{i1},..., x_{iP})'$ the covariates for participant $i$. We set $x_{i1}= 1$ for the intercept and $x_{iP} = g_i$ ($g_i=1$ for treated and 0 for control/placebo in a two-arm trial) for the treatment status. In general, $\mathbf{y}_i$ are partially observed. Let $s_i$ be the dropout pattern according to the time of the last observation. We have $s_i = 0$ for participants with no post-baseline assessment, and $s_i = J$ for completers. Thus, $f(y_{ij}|\mathbf{z}_{ij},\theta_j)$ and $g(y_{ij}|\mathbf{z}_{ij},\theta_j, \tilde{\theta}_j)$ are respectively the distribution of the observed and missing outcomes conditioning on the history.

In the sequential regression approach, the proportional odds model is used for ordinal outcomes with $K$ levels
\begin{equation} \label{eq:clm}
Pr(y_{ij} \leq k|\mathbf{z}_{ij}, \mathbf{\theta}_j) \quad = \quad expit\left(c_{j_k} + \mathbf{\alpha}'_j\mathbf{x}_i + \sum_{t=1}^{j-1}\beta_{jt}y_{it} \right),
\end{equation}
for $k=1,...,K-1$, where $\mathbf{\alpha}_j = (\alpha_{j1},...,\alpha_{jP})'$, $c_{j_1} < ... <c_{j_{K-1}}$, and $expit(\eta) = \exp(\eta)/[1+\exp(\eta)]$. To ensure $c_{j_k} > c_{j_{k-1}}$, the reparameterization $d_{j_k} = \log(c_{j_k}-c_{j_{k-1}})$ is applied, and assign normal prior for $d_{j_k}$:
\begin{align*}
c_{j_1}, d_{j_1}, ...,d_{j_{K-1}} &\uidsim N(\mu_c, \sigma_c^2), \\
c_{j_k} &\sim c_{j_{k-1}} + \exp(d_{j_{k-1}}), k=2,...,K.
\end{align*}
For the binary endpoint, model (\ref{eq:clm}) reduces to
\begin{equation} \label{eq:bin}
Pr(y_{ij} = 1|\mathbf{z}_{ij}, \mathbf{\theta}_j) \quad = \quad expit\left(\mathbf{\alpha}'_j\mathbf{x}_i + \sum_{t=1}^{j-1}\beta_{jt}y_{it} \right),
\end{equation}

\subsection{MI under MAR assumption}
Imputation with MAR assumption is implemented through two algorithmic backends: 
\begin{itemize}
	\item JAGS: It's a gibbs sampler under fully Bayesian setting. Under MAR, $g(y_{ij}|\mathbf{z}_{ij},\mathbf{\theta}_j, \tilde{\mathbf{\theta}}_j)$ is identical to $f(y_{ij}|\mathbf{z}_{ij},\mathbf{\theta}_j)$. Markow chain Monte Carlo (MCMC) with data augmentation (DA) for missing values is realized through \proglang{jags} \citep{Plummer03jags}. Prior distributions have to be specified for all (hyper)parameters. A common prior choice for the regression coefficients is the normal distribution with mean zero and large variance. In \pkg{remiod}, variance parameters are specified as, by default, inverse-gamma distributions. 
\item Algorithm proposed by \cite{Tang2018sim}: The algorithm differentiates intermittent missing data and missingness after dropout. Intermittent missing data is imputed first to make missing pattern become monotone. \cite{Tang2018sim} proposed a Matroplis-Hasting sampler for the monotone data augmentation (MDA) step. The missing data after dropout can be imputed sequentially given the draw of the model parameters and imputed intermittent missing data after the MDA algorithm converges. 
\end{itemize}

\subsection{Controlled MI under MNAR}
\subsubsection{delta-adjusted PMM}
Let $\delta$ be the sensitivity parameter. Then the adjustment is placed onto the parameter corresponding to treatment variable $g_i$:
\begin{equation} \label{eq:clmdelta}
Pr(y_{ij} \leq k|\mathbf{z}_{ij}, \mathbf{\theta}_j) \quad = \quad expit\left[c_{j_k} + \sum_{p=1}^{P-1} \alpha_{jp} x_{ip} + (\alpha_{jP} + \delta) g_i + \sum_{t=1}^{j-1}\beta_{jt}y_{it} \right],
\end{equation}
Model (\ref{eq:clmdelta}) implies MAR in the control group, while the log odds of being in better/worse health status after dropout are reduced compared to those who remain in the trial ($\delta < 0$ if lower scores on $y_{ij}$'s indicate better health. Otherwise, $\delta > 0$). In model (\ref{eq:clmdelta}), it is impossible to estimate $\delta$ from the observed data. Thus, the method of tipping point analysis is often used, that is, we assume $\delta$ is known and repeat the MI inference at a sequence of increasing $\delta$ values to find the tipping point $\delta$, at which the treatment effect becomes insignificant. The MAR-based analysis is said to be robust if the tipping point is large and deemed clinically implausible. 

\subsubsection{Copy reference PMM}
Copy reference (CR) procedure can be presented as follows:
\begin{equation} \label{eq:clmcr}
Pr(y_{ij} \leq k|\mathbf{z}_{ij}, \mathbf{\theta}_j) \quad = \quad expit\left[c_{j_k} + \sum_{p=1}^{P-1} \alpha_{jp} x_{ip} + \sum_{t=1}^{j-1}\beta_{jt}y_{it} \right], \forall j > s_i
\end{equation}
Clearly, under CR, treated participants are assumed to gain no benefit even though the treatment was taken, thus having the same mean response profiles as the reference (i.e.control) participants both before and after dropout \citep{pmid24138436}. 

\subsubsection{Jump to reference PMM}
Jump to reference (J2R) assumes that all treatment benefits are gone immediately after participants discontinue the experimental treatment. The unconditional treatment effect at visit $j (j>s_i)$ that is unadjusted for $(y_{i1}, … , y_{i,j-1})$ needs to be computed for imputing missing values after dropout.

\section{Package structure}
The package \pkg{remiod} has the main function, \fct{remiod}, that performs MCMC sampling. One model must be specified through \code{formula} argument, which is the same as the specification of standard regression models in \proglang{R} and described in detail in Section \ref{sec:modelspec}. Based on the specified model formula and arguments like \code{family} and \code{models}, \pkg{remiod} checks which variables have missingness, determines the order of imputation model sequence, and identifies the variable types (ordinal, binary, categorical, or continuous data) in order to specify appropriate imputation models.

Currently, the package includes two options, defined with the argument \code{algorithm} of \fct{remiod} function, for choosing a backend. For \code{algorithm = "jags"}, \proglang{JAGS} is the backend that performs MCMC sampling \citep{Plummer03jags}. The BUGS language coded model \citep{Lunn2017gy}, related data information, and user-specified settings for the MCMC sampling (see Section \ref{sec:mcset}) are passed to JAGS via the package \pkg{rjags} \citep{Plummer19rjags}. This is an extension of JointAI \citep{erler2021jai}. For \code{algorithm = "tang_seq"}, the sequential modeling approach proposed by \cite{Tang2018sim} is the base of MCMC sampling and imputation. The algorithm is coded with \proglang{R} language.

\fct{remiod} returns MCMC sampled parameters of all imputation models under MAR assumption, as well as all imputed datasets corresponding to sampled parameters in retained MCMC iterations. The MCMC samples under MAR will serve as the source of subsequent controlled MI. \fct{extract\_MIdata} uses the returned object of \textbf{remiod} function to implement other types of controlled imputations and extract an expected number of imputed datasets. 

\begin{Code}
extract_MIdata(object, method="delta", delta=0.1, M=100, minspace=5, mi.setting=NULL)
\end{Code}

The argument \code{method} specifies either MAR or a type of controlled MI approach. Argument \code{M} specifies the number of imputed datasets to be created, with argument \code{minspace} be the minimum number of iterations between iterations eligible for selection. Under default, the seed value specified in \fct{remiod} will be used again, which can be updated through argument \code{mi.setting}, for example, \code{mi.setting=list(seed=1111)}.

\section{Example data}
The Schizophrenia data was based on a randomized controlled trial, i.e. NIMH schizophrenia collaborative study, with 437 patients assigned to either three drugs ($n=329$) or placebo group ($n=108$). The 7-level response of "Severity of Illness" was first analyzed by \cite{pmid3153505}. Later collapsed 4-level ordinal outcome was analyzed by \cite{Gibbons1994}. In this paper, we use the four-level ordered scale. A binary response variable based on the original numerical measurements was also included. The three active drugs was collapsed into one overall drug group, since previous analyses revealed similar effects among the three active drugs. The outcome was measured at weeks 0, 1, \ldots, 6. Most of the observations where made on weeks 0, 1, 3, and 6. The data has six variables:
\begin{itemize}
	\item \code{id}: {patient ID.}
	\item \code{imps79}: {original response measurements on a numerical scale.}
	\item \code{imps79b}: {binary response based on the cut-off value
		of 3.5 to the measurements on a numerical scale: 0 = normal to
		mildly ill and 1 = moderately to extremely ill}
	\item \code{imps79o}: {ordinal response on a 4 category scale,
		1="normal or borderline mentally ill", 2="mildly or moderately ill",
		3="markedly ill", and 4="severely or among the most extremely ill".}
	\item \code{tx}: {treatment indicator: 1 for drug, 0 for placebo.}
	\item \code{week}: {week. Coded from 0 as baseline to week 6.}
\end{itemize}

\begin{figure}[t!]
	\centering
	\includegraphics[scale=0.8]{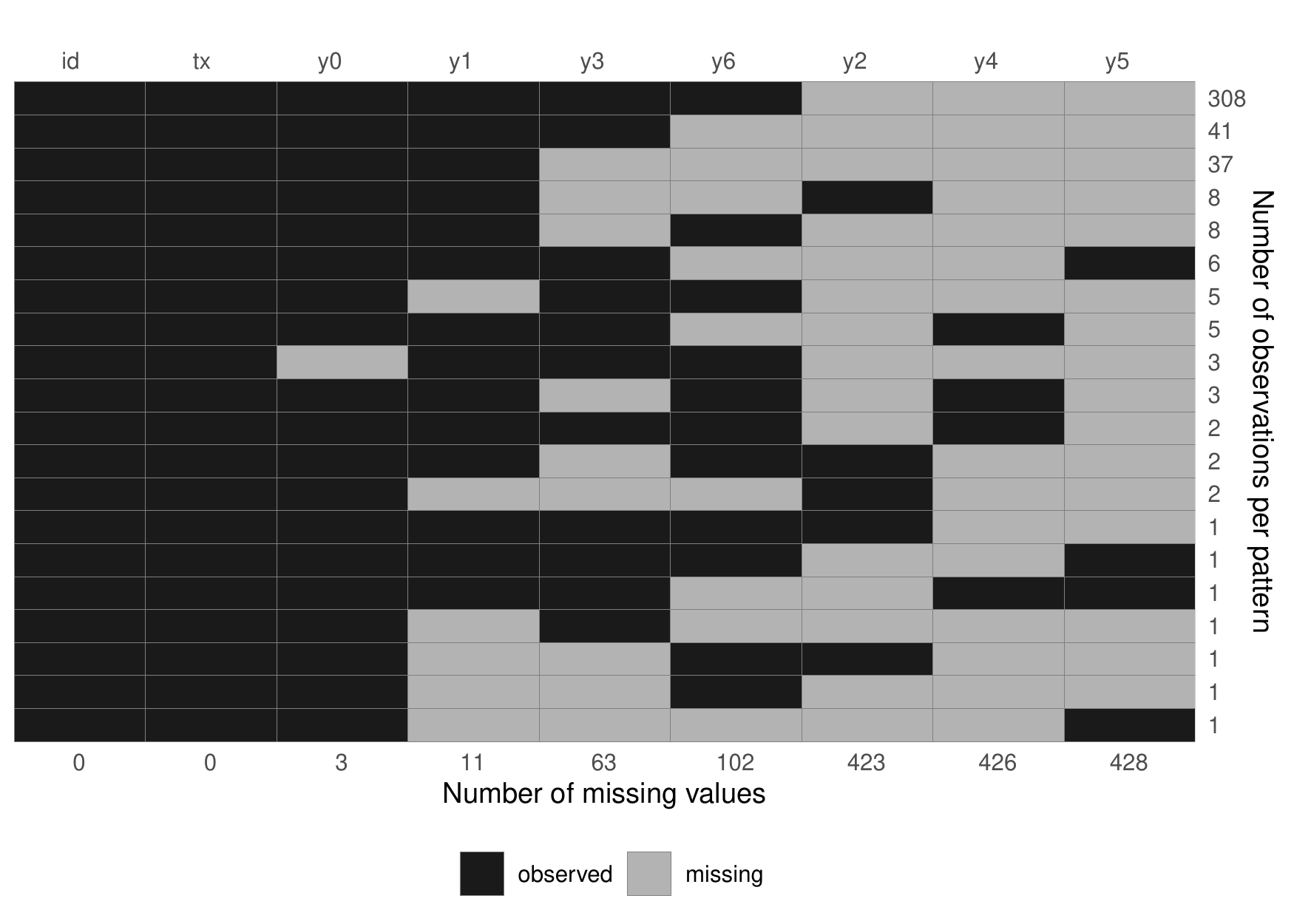}
	\caption{\label{fig:misspat} Pattern of missing values in Schizophrenia data.}
\end{figure}

The original data was in the long format, i.e. the data has a row for observation per subject per visit. Transforming the data into a wide format with prefix \code{y} preceding visit weeks as the variable names of measured outcomes (i.e. \code{schizow} and \code{schizob} data in the package), the missing value pattern of Schizophrenia data can be found in Figure~\ref{fig:misspat}. Note that binary responses need to be convert to factor variables, and ordinal responses to ordered factors, before running \fct{remiod} function.

\section{Specification of models and imputation method} \label{sec:modelspec}

The main analysis function of \fct{remiod} can be called as follow:
\begin{Code}
remiod(formula, data, family = NULL, algorithm = "jags", model_order = NULL,
       models = NULL, method, delta, n.chains = 2, n.iter = 0, thin = 1,...)
\end{Code}
where \code{formula} and \code{data} follows the standard way of specifying regression relationships in \proglang{R}.  Argument \code{family} can set to be \code{NULL} when an ordinal response variable is sent into the function as ordered factors. Otherwise, it is a standard argument for specifying distribution and link function (Table \ref{tab:overview}), which is the same as the one in \code{glm} of R \pkg{stats}. The algorithmic backend can be set up through argument \code{algorithm} . Arguments \code{models} and \code{model_order} are for manually specifying the types of models (e.g. \code{lm} for continuous variables, \code{glm_binomial_logit} for binary variable, etc. See Section 5.5 in \cite{erler2021jai} for more details) and an order for the sequence of models for variables with missingness. 

\subsection{Order of the sequence of imputation models}

When using the default setting of \code{remiod}, the sequence of imputation models are ordered by the number of missing values. Using Schizophrenia data as an example,

\begin{Code}
> testa = remiod(formula = y6 ~ tx + y0 + y1 + y2 + y3 + y4 + y5, 
+                data = schizow, trtvar = 'tx', method = 'MAR', 
+                algorithm = "jags", n.iter = 0, warn = FALSE) 
\end{Code}
\begin{Code}
> list.models(testa)	
Order   model_formula                        
  1     y0 ~ tx                              
  2     y1 ~ tx + y0                         
  3     y3 ~ tx + y0 + y1                    
  4     y2 ~ tx + y0 + y1 + y3               
  5     y4 ~ tx + y0 + y1 + y2 + y3          
  6     y5 ~ tx + y0 + y1 + y2 + y3 + y4     
  7     y6 ~ tx + y0 + y1 + y2 + y3 + y4 + y5
\end{Code}

As Figure \ref{fig:misspat} shows, the number of missing values is 423 for \code{y2}, which is greater than that of \code{y3}. This leads to the imputation model for \code{y3} preceding to the imputation model of \code{y2}. This violates the natural order in visit time. In case the sequence of models is expected to follow the time order of visits, argument \code{model_order = paste0("y",0:5)} can be used to specify the order as \code{y0}, \code{y1}, \code{y2}, \code{y3}, \code{y4}, and \code{y5}, as follows:

\begin{Code}
> testb = remiod(formula = y6 ~ tx + y0 + y1 + y2 + y3 + y4 + y5, 
+                data = schizow, trtvar = 'tx', method = 'MAR', 
+                model_order =paste0("y",0:5), algorithm = "jags",
+                n.iter = 0, warn = FALSE) 
\end{Code}
\begin{Code}
> list.models(testb)	
Order   model_formula                        
  1     y0 ~ tx                              
  2     y1 ~ tx + y0                         
  3     y2 ~ tx + y0 + y1                    
  4     y3 ~ tx + y0 + y1 + y2               
  5     y4 ~ tx + y0 + y1 + y2 + y3          
  6     y5 ~ tx + y0 + y1 + y2 + y3 + y4     
  7     y6 ~ tx + y0 + y1 + y2 + y3 + y4 + y5
\end{Code}

\subsection{Ordinal variable as covariate}
In the case of imputing ordinal responses, ordinal measurements in previous visits will be used as covariates to impute missing measurements at subsequent visits (see the example model lists in the above section). In this scenario, the ordinal covariates can be treated as either categorical variables or continuous variables. Argument \code{ord_cov_dummy} can be set to be \code{TRUE} if the user want to treat an ordinal variable as categorical one (a series of dummy variables will be created based on specified reference category). Otherwise, the ordinal covariates will be treated as continuous variables.

\subsection{MI Method}
The major objective of this package is to generate completed datasets using MI, so \code{method} argument is required to have an input. Default is MAR. Currently, the available options include \code{MAR}, \code{J2R}, \code{CR}, and \code{delta}. For \code{method = "delta"}, argument \code{delta} should follow to specify a numerical values used in delta adjustment (see Equation \ref{eq:clmdelta} for details).

\begin{table}[t!]
\caption{\label{tab:overview}  Possible choices for the model family and link functions in GLM models.}
\centering
\begin{tabular}{lp{7.4cm}}
\hline
Distribution    & Link \\ \hline
gaussian        & identity, log, inverse \\ 
binomial        & logit, probit, log, cloglog \\ 
gamma           & inverse, identity, log \\ 
poisson         & log, identity \\ \hline
\end{tabular}
\end{table}

\section{Inherited settings} \label{sec:mcset}

Arguments for specifying settings of the MCMC sampling in \pkg{JointAI}, including \code{n.chains}, \code{n.iter}, \code{n.adapt}, \code{thin}, and \code{inits} are inherited. The rules to set up MCMC sampling described in \cite{erler2021jai} are applicable in \fct{remiod}.

\section{Illustrations} \label{sec:illustrations}

\begin{figure}[t!]
	\centering
	\begin{subfigure}{1\textwidth}
		\caption{Tang's algorithm}
		\includegraphics[scale=0.6]{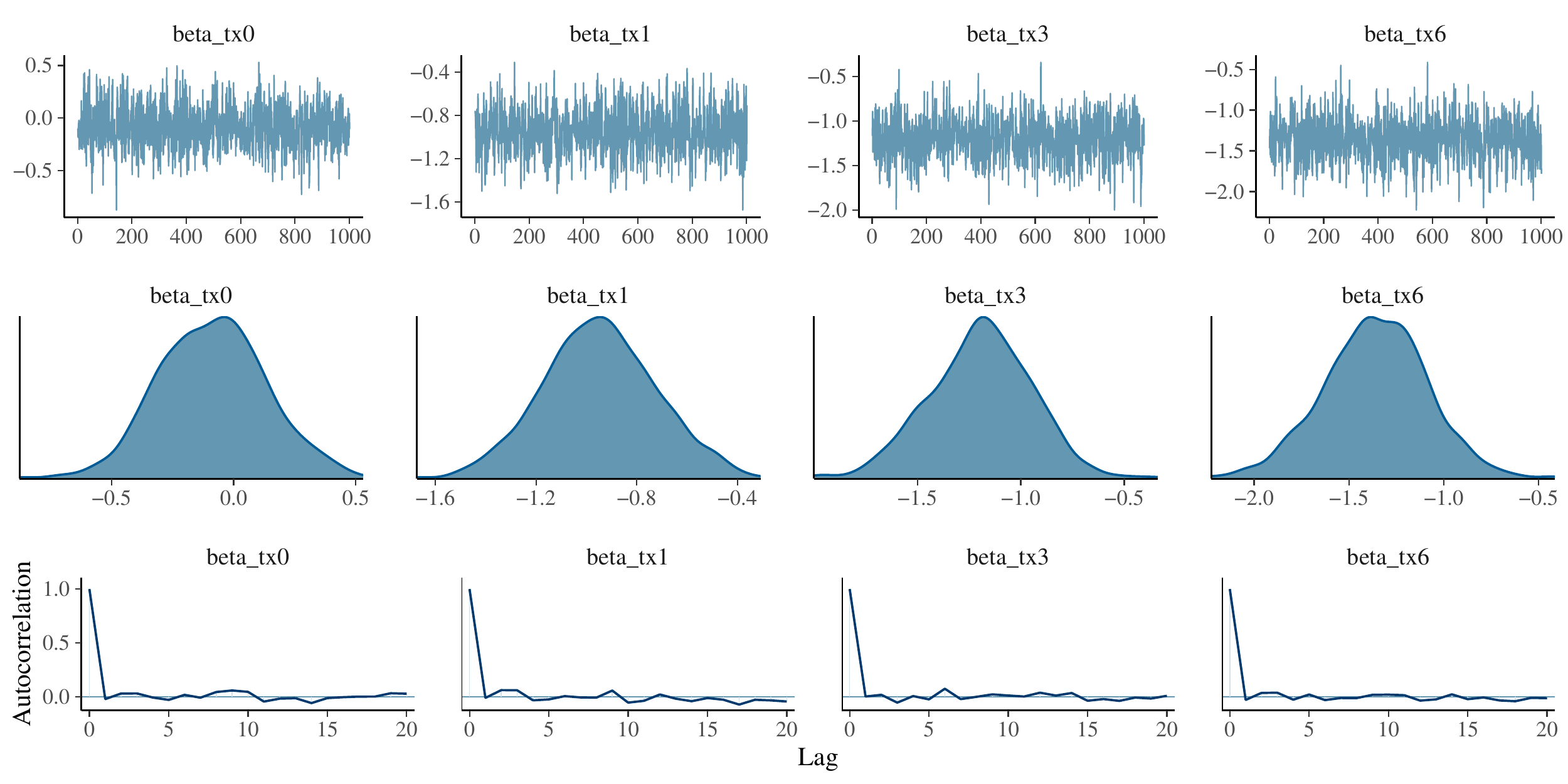}
		\label{fig:tracetang}
	\end{subfigure}
	\newline
	\begin{subfigure}{1\textwidth}
		\caption{JAGS algorithm.}
		\includegraphics[scale=0.6]{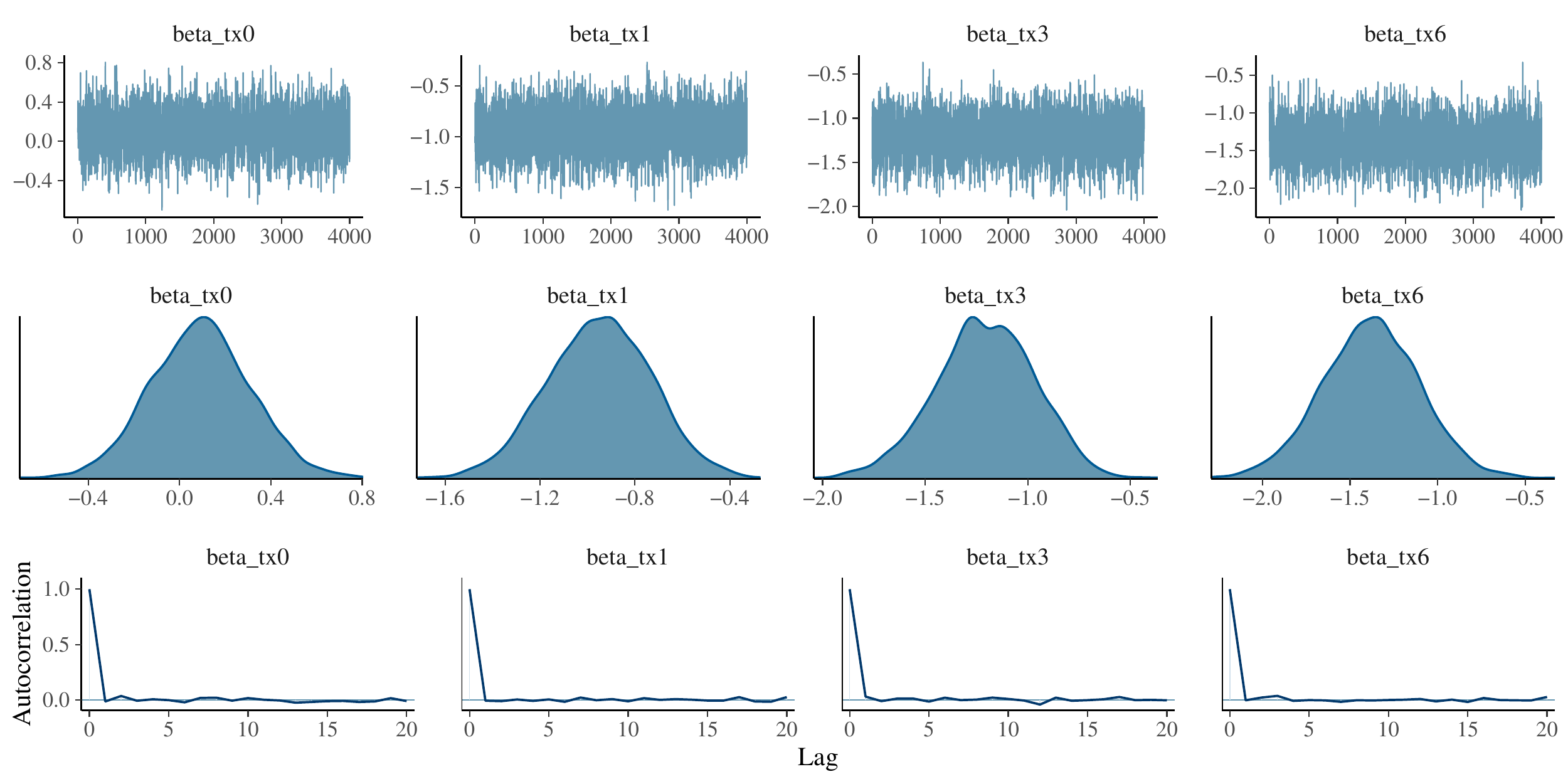}
		\label{fig:tracejags}
	\end{subfigure}
	\caption{Trace, density, and ACF plots of sampled coefficients for treatment variable in imputation models. Number 0, 1, 3, and 6 in the names of coefficient indicates visit at baseline and weeks 1, 3, and 6.}
	\label{fig:traces}
\end{figure}

We analyze the ordinal response of the NIMH schizophrenia collaborative study with multiple imputation based on the algorithm inherited in \code{JAGS} and the method proposed by \cite{Tang2018sim}. In order to see whether we can replicate the results obtained by \cite{Tang2018sim}, the same data used by Tang, i.e. data collected at baseline and weeks 1, 3, and 6, are considered in this analysis. Trace, density, and ACF plots of sampled coefficients for treatment variable in imputation models are shown in Figure \ref{fig:traces}.

M = 1000 datasets are imputed from every 100th iteration after a burn-in period of 10,000 iterations. Each imputed dataset is analyzed using the cumulative logistic regression by visit. The results are summarized in Table \ref{tabresult}. Estimated treatment effects under all scenario are matched, considering the variation of Bayesian sampling, to those presented in \cite{Tang2018sim}. It also indicates that the two algorithmic backends perform great comparability.

\captionsetup[sub]{justification=justified,singlelinecheck=false}

\begin{table}
	\caption{Estimated treatment effects by visit and associated Rubin's variance 
		for the National Institute of Mental Health (NIMH) schizophrenia trial}
	\label{tabresult}
	\centering
	\begin{tabular}[t]{ccccccccc}
		\toprule
		\multicolumn{1}{c}{ } & \multicolumn{1}{c}{ } & \multicolumn{3}{c}{JAGS} & \multicolumn{1}{c}{} & \multicolumn{3}{c}{\cite{Tang2018sim}} \\
		\cmidrule(l{3pt}r{3pt}){3-5} \cmidrule(l{3pt}r{3pt}){7-9}
		
		\multicolumn{1}{c}{ } & \multicolumn{1}{c}{ } & \multicolumn{1}{c}{MI } & \multicolumn{2}{c}{Rubin's Variance} & \multicolumn{1}{c}{} & \multicolumn{1}{c}{MI} & \multicolumn{2}{c}{Rubin's Variance} \\
		\cmidrule(l{3pt}r{3pt}){4-5} \cmidrule(l{3pt}r{3pt}){8-9}
		
		Method & Visit & Estimate & Between & Within & & Estimate & Between & Within\\
		\midrule
		MAR & 1 & 0.790 & 0.001 & 0.044 &  & 0.790 & 0.001 & 0.044\\
		 & 2 & 1.331 & 0.005 & 0.045 &  & 1.333 & 0.005 & 0.045\\
		 & 3 & 1.847 & 0.014 & 0.049 &  & 1.843 & 0.013 & 0.049\\
		\addlinespace
		J2R & 1 & 0.783 & 0.001 & 0.044 &  &  &  & \\
		 & 2 & 1.221 & 0.014 & 0.044 &  &  &  & \\
		 & 3 & 1.459 & 0.042 & 0.045 &  &  &  & \\
		\addlinespace
		CR & 1 & 0.784 & 0.001 & 0.044 &  & 0.785 & 0.001 & 0.044\\
		 & 2 & 1.248 & 0.008 & 0.044 &  & 1.242 & 0.006 & 0.044\\
		 & 3 & 1.597 & 0.017 & 0.046 &  & 1.598 & 0.014 & 0.046\\
		\addlinespace		
		Delta & 1 & 0.784 & 0.001 & 0.044 &  & 0.785 & 0.001 & 0.044\\
		 & 2 & 1.259 & 0.008 & 0.044 &  & 1.257 & 0.006 & 0.044\\
		 & 3 & 1.660 & 0.020 & 0.047 &  & 1.655 & 0.014 & 0.047\\
		\bottomrule
	\end{tabular}
\end{table}


\section{Discussion} \label{sec:summary}

This paper has presented the first comprehensive software package for reference-based multiple imputation methods based on cumulative logistic model and generalized linear models. The package implements two different MCMC algorithms as the backends of imputing missing data, and includes functionality for three different types of controlled imputation based on PMM. The package includes functions for extracting imputed data sets and visualizations of the results after running \fct{remiod}. It also includes a built-in example to illustrate the functionality of the package.

Although the original objective to develop the package is for analyzing binary and ordinal variables with missing values, the JAGS-based sampler can actually be applied to all GLM modelable data, including data generated from normal distribution, Poisson distribution, and Gamma distribution (Table \ref{tab:overview}). PMM-based controlled multiple imputation can thus be applicable to analyze these types of data.

Provided that the measurement at each visit is generated from a proper univariate probability model, a sequential specification always defines a coherent joint model based on the factorization theorem. However, different orderings will generally lead to different joint distributions and potentially different fits. Heuristics have been proposed for selecting the order, for example ordering variables by their types \citep{Ibrahim1999MissingCI} or percentage of missing values \citep{robin1990asa}. The latter is particularly well-motivated when the missing data are monotone. Package \pkg{remiod} allows manually ordering sequential models, extending the flexibility in application, especially in the scenario where natural ordering of sequential models is more intuitive.  

\pkg{remiod} provides a valid and user-friendly way to implement controlled multiple imputation. To expand the range of \pkg{remiod} settings, several extensions are planned, including
\begin{itemize}
	\item Implementation of additional model types, for example, univariate and multivariate probit model based imputation.
	\item implementation of other types of controlled imputation, e.g. copy increment in reference, extended/modified copy reference \citep{Tang2016AnEM}, etc.
	\item unbiased variance estimator for controlled multiple imputation.
\end{itemize}

\newpage
\bibliographystyle{unsrtnat}
\bibliography{refs}

\end{document}